\documentclass[aps,twocolumn,prb,floatfix]{revtex4-1}  
\usepackage{graphicx}
\usepackage{multirow}
\usepackage{array}
\usepackage[usenames]{color}
\usepackage{amsmath}
\usepackage{amssymb}
\usepackage{ulem}
\begin{document}

\title{Accelerating the convergence of path integral dynamics with a generalized Langevin equation}
\author{Michele Ceriotti}
\email{michele.ceriotti@phys.chem.ethz.ch}
\affiliation{Computational Science, Department of Chemistry and Applied Biosciences, ETH Z\"urich, USI Campus, Via Giuseppe Buffi 13, CH-6900 Lugano, Switzerland}

\author{David E. Manolopoulos}
\affiliation{Physical and Theoretical Chemistry Laboratory, Oxford
University, South Parks Road, Oxford OX1 3QZ, UK}

 \author{Michele Parrinello}
\affiliation{Computational Science, Department of Chemistry and Applied Biosciences, ETH Z\"urich, USI Campus, Via Giuseppe Buffi 13, CH-6900 Lugano, Switzerland}

\begin{abstract}
The quantum nature of nuclei plays an important role in the accurate modelling of light atoms such as hydrogen, but it is often neglected in simulations due to the high computational overhead involved. It has recently been shown that zero-point energy effects can be included comparatively cheaply in simulations of harmonic and quasi-harmonic systems by augmenting classical molecular dynamics with a generalized Langevin equation (GLE). Here we describe how a similar approach can be used to accelerate the convergence of path integral (PI) molecular dynamics to the exact quantum mechanical result in more strongly anharmonic systems exhibiting both zero point energy and tunnelling effects. The resulting PI-GLE method is illustrated with applications to a double-well tunnelling problem and to liquid water.
\end{abstract}
\maketitle

\section{Introduction}

Atomistic computer simulations have become an important complement to experimental 
measurements in studies of a wide variety of physical, chemical and biological systems. 
However, in order to reduce computational effort that is required for larger systems, 
a number of approximations are typically made in these simulations. 
One that is commonly employed is to assume that the atomic nuclei behave as classical 
particles, even when the interactions between them are obtained from an {\em ab initio} 
calculation. 
This is a reasonable assumption for heavy atoms at high temperatures. 
But for lighter atoms -- and hydrogen in particular -- significant deviations are to be 
expected from classical behavior even at room temperature. 

When empirical interaction potentials are used to compute the forces acting on the nuclei, 
nuclear quantum effects are often implicitly accounted for, because force fields are 
typically parameterised so as to agree with experimental data when used in classical 
molecular dynamics simulations. 
When {\em ab initio} methods are used to describe the interactions between the nuclei, 
there is no such parameterisation for nuclear quantum effects, and the error that results 
from assuming purely classical behavior of the nuclear motion is often comparable to that 
stemming from an approximate treatment of electron correlation.\cite{Morrone08}

The conventional approach to including nuclear quantum effects exploits the isomorphism 
between the quantum mechanical partition function of the physical system and the classical 
partition function of an extended problem consisting of a necklace (or ring polymer) of 
replicas of the system in which corresponding atoms are connected by harmonic springs.\cite{Chandler81} 
As the number of replicas (or beads of the necklace) is increased, this imaginary time 
path integral\cite{Feynman65} (PI) method samples an ensemble that converges systematically 
to that of the quantum problem with distinguishable particles, at the expense of a computational effort that increases 
in proportion to the number of beads. 

Methods have been devised to reduce this effort by splitting the calculation of forces 
into an inexpensive, short-ranged contribution that is computed on every bead, and a 
long-range contribution that is evaluated on a contracted ring polymer with fewer beads.\cite{Markland08b,Markland08c,Fanourgakis09} 
These methods are straightforward to implement in simulations with empirical force fields. 
However they cannot presently be used in {\em ab initio} simulations, where the splitting 
of the forces into an inexpensive short-range contribution plus a more expensive long-range 
contribution is significantly more difficult to arrange. 
Approximate quantum methods such as the Feynman-Hibbs\cite{Feynman65,Voth91} approach are 
also difficult to use in {\em ab initio} simulations, because they require the computation of the Hessian.

A more general approach to reducing the effort of path integral simulations is therefore 
called for, and there are certain indications in the literature that this might now be possible. 
In particular, it has recently been shown that colored noise, generalized Langevin equation 
(GLE) thermostats can be used not only to enhance the sampling of classical and path integral 
molecular dynamics,\cite{Ceriotti10a,Ceriotti10b} but also to modify conventional molecular 
dynamics so as to include nuclear quantum effects in mildly anharmonic systems in which 
zero-point energy plays a significant role.\cite{Ceriotti09b} 
This approach involves negligible overhead with respect to purely classical dynamics, provides 
very naturally the proton momentum distribution (which is relevant for comparison with inelastic 
neutron scattering experiments), and can be applied with the same ease to empirical 
and {\em ab initio} force fields. However, it does not provide a satisfactory description of 
more subtle quantum effects such as tunnelling, and its accuracy cannot be systematically improved.

In this paper, we will show that it is possible to combine a GLE thermostat with path integral 
molecular dynamics (PIMD) in such a way as to exploit the best features of both techniques. 
In particular, by tuning the properties of the correlated noise in an appropriate GLE, we will 
show that the systematic convergence of PIMD to the exact quantum mechanical result can be greatly 
accelerated, leading to significant computational savings for a given level of accuracy. 
The resulting PI+GLE scheme provides an equally reliable description of zero point energy and 
tunnelling effects, it is equally applicable to simulations with empirical and {\em ab initio} 
force fields, and unlike the original single-bead ``quantum thermostat'' discussed above it can 
be systematically improved simply by increasing the number of path integral beads.

The outline of the paper is as follows. In Section~II we briefly recall a few concepts from 
path integral and generalized Langevin equation methods, and describe our strategy to have 
them work in synergy. In Section~III we present a systematic study of a one-dimensional 
quartic double well potential, discussing the effect of zero point energy and tunnelling on 
the convergence of our method. In Section~IV we examine how the method performs for liquid water, 
and in Section~V we draw our conclusions.

\section{Combining path integrals with the generalized Langevin equation}
\subsection{Path integral methods}

Let us first recall the basic principles of imaginary time path integral methods, so as to 
introduce our notation. We will only discuss the one-dimensional case and refer the reader to the 
literature\cite{Feynman65,Barker79,Chandler81,Parrinello82,Parrinello84,Ceperley95,Schulman05} 
for a more detailed discussion.

Consider the Hamiltonian for a particle in a one-dimensional potential, 
\begin{equation}
\hat{H}=\frac{1}{2}\hat{p}^2+V(\hat{q}),\label{eq:q-ham}
\end{equation}
where $\hat{p}$ and $\hat{q}$ are the mass-scaled momentum and position operators and the 
potential $V(q)$ is assumed to be such that the quantum mechanical partition function 
\begin{equation}
Z=\mathrm{tr}\bigl[e^{-\hat{H}/k_B T}\bigr] \label{eq:q-z}
\end{equation}
is well defined. The path integral formalism avoids the solution of the Schr{\"o}dinger 
equation for the Hamiltonian in Eq.~(\ref{eq:q-ham}) and allows one instead to sample configurations 
consistent with the quantum mechanical equilibrium distribution by exploiting an isomorphism 
with an extended classical problem.\cite{Chandler81} 
Indeed a standard Trotter-product\cite{Schulman05} approximation to the Boltzmann operator in 
Eq.~(\ref{eq:q-z}) yields the following extended phase space expression for the partition function
\begin{equation}
Z\approx Z_P=\frac{1}{(2\pi\hbar)^P} \int \mathrm{d}^P \mathbf{p}
\int \mathrm{d}^P \mathbf{q}\, e^{-H_P(\mathbf{p},\mathbf{q})/Pk_B T},\label{eq:pi-z}
\end{equation}
where $H_P(\mathbf{p},\mathbf{q})$ is the classical Hamiltonian of a fictitious ring polymer 
composed of $P$ replicas of the system connected by harmonic springs
\begin{equation}
H_P(\mathbf{p},\mathbf{q})=\sum_{j=0}^{P-1}\left[\frac{1}{2}p_j^2+
  V(q_j)+\frac{1}{2}\omega_P^2 (q_{j}-q_{j+1})^2\right],\label{eq:ham-p}
\end{equation}
with $\omega_P=Pk_BT/\hbar$ and $q_P\equiv q_0$. The error in this approximation is $O(P^{-2})$ 
and so vanishes to leave the exact quantum mechanical result in the limit as $P\to\infty$.\cite{Schulman05}

The momenta are often integrated out of Eq.~(\ref{eq:pi-z}) to leave a purely configurational integral 
which forms the basis of the path integral Monte Carlo (PIMC) technique.\cite{Barker79} 
One can however retain the momenta, and describe the dynamical evolution of the ring 
polymer by means of Hamilton's equations. This is the PIMD approach,\cite{Parrinello82,Parrinello84} 
which provides a particularly efficient way to sample the configuration space in situations such 
as molecular liquid simulations in which an effective PIMC calculation would require 
complicated collective moves. A fully converged PIMD calculation produces the exact 
thermodynamic and structural properties of the quantum mechanical system. 
Although it is something of an aside to the present work, in which we shall be concerned 
exclusively with static equilibrium properties, it is also now well established that the centroid 
molecular dynamics\cite{Cao94,Jang99} and ring polymer molecular dynamics\cite{Craig04,Braams06} 
generalizations of PIMD can be used to provide quite reasonable estimates of dynamical 
properties such as diffusion coefficients and chemical reaction rates.

The extension to three dimensions and to an arbitrary number of interacting distinguishable 
particles is straightforward. But rather than describe this extension here, let us consider 
instead the simple case of a one-dimensional harmonic potential, $V(q)=\omega^2q^2/2$. 
The result will be used in what follows. For this simple model, the integral in Eq.~(\ref{eq:pi-z}) can be 
evaluated by transforming to the $P$ normal modes $\{q_k\}_0^{P-1}$ of the ring polymer with frequencies
\begin{equation}
\omega_k=\sqrt{\omega^2+4 \omega_P^2 \sin^2 (k\pi/P)}.\label{eq:omegak}
\end{equation}
It is then easy to show that the equilibrium position distribution of each bead of the ring polymer 
will be a Gaussian centred on $q=0$, with a variance 
\begin{equation}
\left<q^2\right>_P
={1\over P}\sum_{k=0}^{P-1}\left<q_k^2\right>=k_B T\sum_{k=0}^{P-1} \frac{1}{\omega_k^2},
\end{equation}
which converges to the correct quantum mechanical thermal expectation value 
\begin{equation}
\left<q^2\right> = {\hbar\over 2\omega} \coth {\hbar\omega\over 2k_BT} \label{eq:pi-q2}
\end{equation}
in the limit as $P\to\infty$.

\subsection{Generalized Langevin equations}

It has recently been shown that an appropriate generalized Langevin equation (GLE) thermostat 
can be used to sample configurations for a harmonic oscillator that are consistent with the 
quantum mechanical variance in Eq.~(\ref{eq:pi-q2}) using just a single replica of the system, thereby avoiding 
the overhead of a $P$-bead PI simulation.\cite{Ceriotti09b}

The basic idea behind GLE thermostats is to construct a linear, Markovian stochastic differential 
equation (SDE) in an extended {\em momentum} space which is coupled to the Hamiltonian dynamics 
in such a way that the equations of motion read
\begin{equation}
\begin{split}
  \dot{q}=&\ p\\
\!\left(\! \begin{array}{c}\dot{p}\\ \dot{\mathbf{s}} \end{array}\!\right)\!=&
\left(\!\begin{array}{c}-V'(q)\\ \mathbf{0}\end{array}\!\!\right)
\!-\!\left(\!
\begin{array}{cc}
a_{pp} & \mathbf{a}_p^T \\ 
\bar{\mathbf{a}}_p & \mathbf{A}
\end{array}\!\right)\!
\left(\!\begin{array}{c}p\\ \mathbf{s}\end{array}\!\right)\!+\!
\left(\!
\begin{array}{cc}
b_{pp} & \mathbf{b}_p^T \\ 
\bar{\mathbf{b}}_p & \mathbf{B}
\end{array}\!\right)\!
\left(\!\begin{array}{c}\multirow{2}{*}{$\boldsymbol{\xi}$}\\ \\\end{array}\!\right),
\end{split}
\label{eq:mark-sde}
\end{equation}
where $\boldsymbol{\xi}$ is a vector of $n+1$ uncorrelated Gaussian random numbers with
$\left<\xi_i\left(t\right)\xi_j\left(0\right)\right>=\delta_{ij}\delta\left(t\right)$. 
It can be shown that the resulting trajectories are statistically equivalent to those obtained 
from a non-Markovian Langevin equation involving only $p$ and $q$,\cite{Zwanzig01}
\begin{equation}
\begin{split}
 \dot{q}=&\ p \\
 \dot{p}=&-V'(q)-\int_{-\infty}^t K(t-s) p(s)\mathrm{d} s +\zeta(t),
\end{split}
\label{eq:nonmark-sde}
\end{equation}
where the friction kernel $K(t)$ and the noise correlation function $H(t)=\left<\zeta(t)\zeta(0)\right>$ 
are related by analytical expressions to the matrices that appear in Eq.~(\ref{eq:mark-sde}). 
These relationships, together with a more detailed discussion, can be found in Ref.~\cite{Ceriotti10a}

The only non-linear term in Eq.~(\ref{eq:mark-sde}) is the force $V'(q)$. If the potential is harmonic, the force 
depends linearly on $q$, and the whole set of equations describe an Ornstein-Uhlenbeck process,\cite{Gardiner03} 
which can be solved analytically to yield closed-form expressions for static and dynamic properties 
of the trajectory. Based on these expressions, one can iteratively refine the parameters $a_p$, ${\bf a}_p^T$, 
$\overline{\bf a}_p$, ${\bf A}$, $b_p$, ${\bf b}_p^T$, $\overline{\bf b}_p$ and ${\bf B}$ in Eq.~(\ref{eq:mark-sde}) subject 
to certain positivity constraints until the desired response of the thermostat is obtained.\cite{Ceriotti10a}

When considering the generalisation of this strategy to a multi-dimensional harmonic problem, one realises 
that, because of the linear nature of the SDE and of the consequent rotational invariance, the dynamics 
will conform to the analytical predictions obtained from the Ornstein-Uhlenbeck process even if the thermostats 
are applied to Cartesian coordinates, without the need to diagonalise the Hessian and transform to normal modes. 
Moreover, it has been demonstrated in previous work that the analytical properties of the thermostat 
obtained in the harmonic limit provide a meaningful prediction of its behavior for anharmonic problems 
with a similar range of frequencies.\cite{Ceriotti10a}

By employing this strategy it is possible to obtain a number of useful effects, such as efficient sampling 
of the canonical distribution in constant-temperature MD. In this application, a fluctuation-dissipation theorem 
must hold, which relates the friction kernel and the memory of the noise in Eq.~(\ref{eq:nonmark-sde}) through $H(t)=k_B T K(t)$. 
More generally, when this condition is not imposed, one obtains a non-equilibrium dynamics in which a 
frequency-dependent effective temperature is enforced. 
Either way, one can calculate the stationary covariance of the harmonic dynamics in the 
$(q,p,\mathbf{s})$ extended phase space, and the mean squared fluctuations of the positions and momenta, 
which we will label $c_{qq}(\omega)$ and $c_{pp}(\omega)$, respectively. It is then possible to design a 
GLE dynamics which behaves as a ``quantum  thermostat'';\cite{Ceriotti09b} i.e., which enforces probability 
distributions of positions and momenta that are consistent with those expected for a quantum harmonic oscillator, 
\begin{equation}
c_{qq}(\omega)=\frac{1}{\omega^2}c_{pp}(\omega)=\frac{\hbar}{2\omega}\coth \frac{\hbar\omega}{2k_BT},\label{eq:cppcqq}
\end{equation}
and does so over a wide range of frequencies.

When applying this idea to a multi-dimensional system one faces the problem of zero-point energy leakage.\cite{Habershon09b} 
Anharmonic coupling causes a flow of heat from high-frequency to low-frequency vibrations, and a departure 
from the desired behavior in Eq.~(\ref{eq:cppcqq}). This problem can be mitigated to some extent by exploiting 
the flexibility of the GLE in Eq.~(\ref{eq:mark-sde}) to enhance the coupling strength of the thermostat and ensure 
that all vibrations are maintained at the correct effective temperature. 
This approach has been shown to give satisfactory results in a number of realistic, condensed-matter applications, 
ranging from the calculation of diamond-graphite coexistence curves\cite{Khaliullin10} to the proton 
momentum distribution in hydrogen-storage materials.\cite{Ceriotti10c} 

Being simple to implement and computationally inexpensive, this ``quantum thermostat" is an important step 
towards a routine treatment of nuclear quantum effects in molecular dynamics. It lacks however two desirable 
features; namely, the possibility of treating more subtle quantum effects such as tunnelling, and the 
possibility of increasing the accuracy in a systematic way. Since both these requirements are met by PI 
methods, one suspects that it might be possible to develop a synergistic approach which combines path integrals 
with the GLE thermostat so as to obtain accurate results without the effort of a fully-converged PI simulation. 
The implementation of such a PI+GLE approach is the subject of the next section.

\subsection{A synergistic combination}

When a GLE thermostat is applied to a path integral molecular dynamics simulation, the internal frequencies of the 
ring polymer necklace will be present along with the physical vibrations of the system. 
It is therefore once again instructive to consider a harmonic model with frequency $\omega$, for which both the 
ring polymer frequencies and the stochastic dynamics can be treated analytically. 

In order to construct a non-equilibrium Langevin dynamics that enforces the quantum mechanical equilibrium 
distribution corresponding to the frequency-dependent fluctuations in Eq.~(\ref{eq:cppcqq}), one must allow in the path integral 
context for the fact that the average of $q^2$ is obtained from a sum over the ring polymer normal modes,
\begin{equation}
 \left<q^2\right>_P=\frac{1}{P}\sum_{k=0}^{P-1} \left<q_k^2\right>=\frac{1}{P}\sum_{k=0}^{P-1} c_{qq}(\omega_k),\label{eq:qp}
\end{equation}
where the normal mode frequencies $\omega_k$ are given in Eq.~(\ref{eq:omegak}). Here the last equality holds if a GLE which results in the frequency-dependent position fluctuation $c_{qq}(\omega)$ has been applied separately to each bead of the ring polymer.

It follows from this that one cannot simply tune the parameters in the GLE acting on each bead of the necklace so that $c_{qq}(\omega)$ 
is given by Eq.~(\ref{eq:cppcqq}). Instead, the appropriate frequency-dependent distribution to be enforced depends on the 
number of beads in the necklace, and can be obtained by solving
\begin{equation}
\frac{1}{P}\sum_{k=0}^{P-1} c_{qq}(\omega_k)=\frac{\hbar}{2 \omega} \coth\frac{\hbar \omega}{2 k_B T}.
\label{eq:cqq-equation}
\end{equation}
The first task is therefore to solve this functional equation for $c_{qq}(\omega)$, recalling again 
that the frequencies $\omega_k$ are related to the frequency $\omega$ of the harmonic oscillator by Eq.~(\ref{eq:omegak}); 
the solution will be a universal function of $\omega$ that is equally applicable to any harmonic oscillator, 
just as in the case of the original quantum thermostat\cite{Ceriotti09b} that enforces the condition on $c_{qq}(\omega)$ in Eq.~(\ref{eq:cppcqq}).

\begin{figure}[t]
\includegraphics{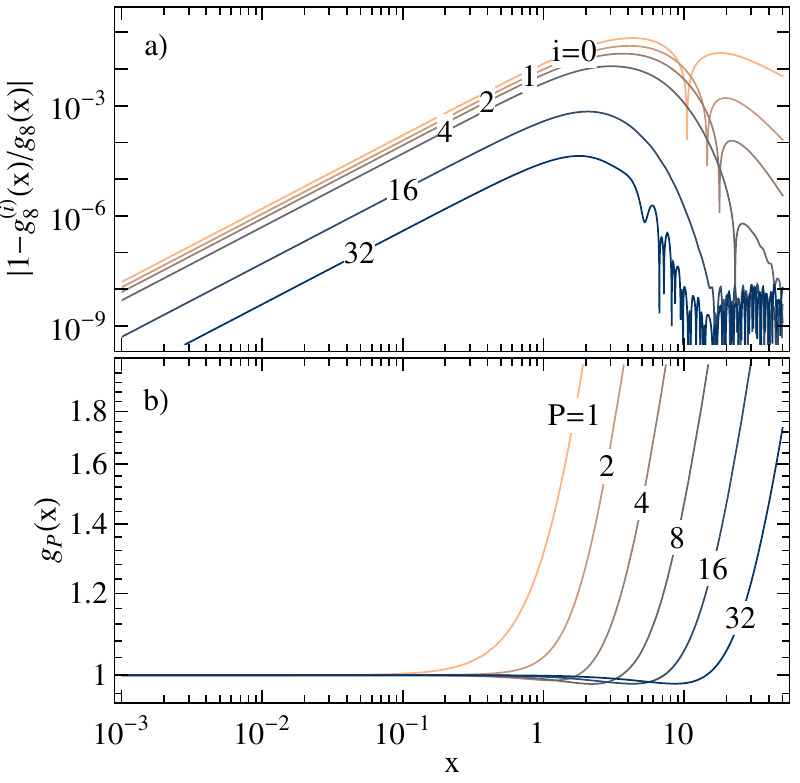}
\caption{(a) Relative error in the estimates of $g_8(x)$ at different iterations of the self-consistent procedure in Eq.~(\ref{eq:gp-iterative}), with $\alpha=1/8$. The wiggles which appear after many iterations occur because each $g_P^{(i)}(x)$ is approximated as a spline interpolation before computing $g_P^{(i+1)}(x)$. These wiggles can be systematically reduced by using a denser grid in $x$. (b) Converged $g_P(x)$ curves for different bead numbers. Note that the curves with larger values of $P$ are flat up to larger values of $x$. }
\end{figure}

\begin{figure}[t]
\includegraphics{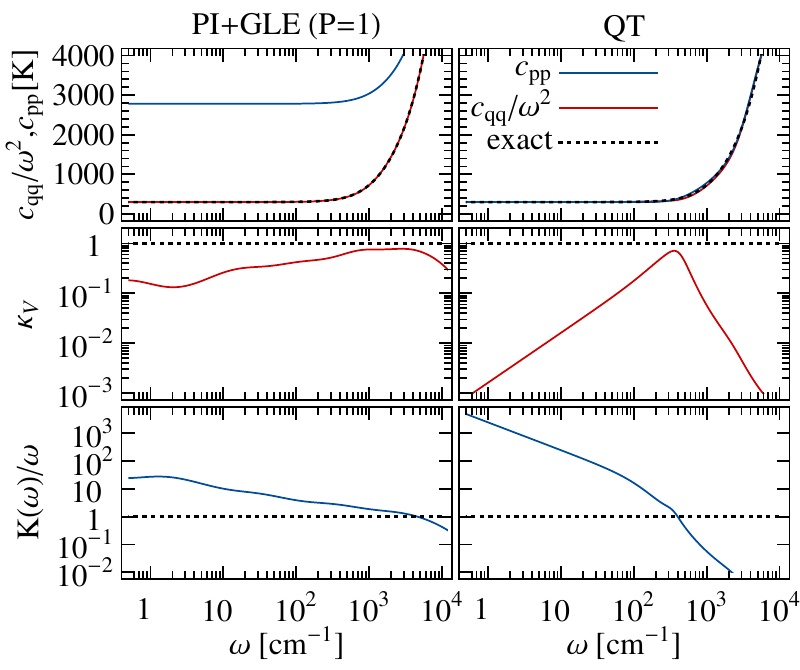}
\caption{Comparison between the properties of the stochastic dynamics which result from the fitting strategy used in the present work (left column) and that used for the original quantum thermostat\cite{Ceriotti09b} (right column), for simulations at $T=298$~K. Note that by sacrificing the fit to the fluctuations of momenta $[c_{pp}(\omega)]$ it is possible to obtain a perfect fit to $c_{qq}(\omega)$ and a better sampling efficiency $\kappa_V(\omega)=1/[\tau_V(\omega)\omega]$, where $\tau_V(\omega)$ is the correlation time of the potential energy for a harmonic oscillator with frequency $\omega$. In the case of PI+GLE it is less essential to enforce strong coupling to avoid zero point energy leakage, and it is therefore possible to avoid the overdamping which hinders the sampling in the case of the original quantum thermostat. This overdamping is clear from the bottom right-hand panel, which shows that $K(\omega)\gg \omega$ at low frequencies where $K(\omega)$ is the Fourier transform of the memory kernel in Eq.~(\ref{eq:nonmark-sde}). }
\end{figure}

Before we describe our approach to solving Eq.~(\ref{eq:cqq-equation}), let us transform it 
into dimensionless form by defining $x=\hbar\omega/2k_B T$, $h(x)=x\coth x$, 
and $g_P(x)=(k_BT/P)(2x/\hbar)^2c_{qq}(2xk_BT/\hbar)$. With these definitions, Eq.~(\ref{eq:cqq-equation}) becomes
\begin{equation}
\sum_{k=0}^{P-1} \frac{g_P(x_k)}{x_k^2/x^2} = h(x), \label{eq:h-equation}
\end{equation}
where
\begin{equation} 
\quad x_k^2=x^2+P^2\sin^2 \frac{k\pi}{P}. \label{eq:xk}
\end{equation}
The solution to Eq.~(\ref{eq:h-equation}) is not unique, and one must pick a particular solution by means 
of appropriate boundary conditions. In particular, one would like to enforce a physical behavior on the solution, 
with no discontinuities and reasonable asymptotic forms in the $x\rightarrow 0$ and $x\rightarrow \infty$ limits. 
The tentative solution 
\begin{equation}
g^{(0)}_P(x)=h(x/P)\label{eq:gp-zero}
\end{equation}
satisfies these requirements, and it is a good approximation to $g_P(x)$ from several points of view. 
First of all, it is the exact solution in the one-bead case (which corresponds to the bare quantum 
thermostat) and in the infinite bead limit, where it yields a constant effective temperature on all 
frequencies (which is correct, as in this case PIMD alone is sufficient to converge to the 
appropriate distribution). It also provides the appropriate large-$x$ limit for a smooth 
solution to Eq.~(\ref{eq:h-equation}), for arbitrary $P$.

In order to refine this tentative solution, one can cast Eq.~(\ref{eq:h-equation}) into a fixed-point 
iteration, by singling out the $k=0$ term:
\begin{equation}
g_P(x)=h(x) -\sum_{k=1}^{P-1} \frac{g_P(x_k)}{x_k^2/x^2}. 
\label{eq:gp-fixedpoint}
\end{equation}
We have found empirically that a self-consistent iterative procedure that yields an exact 
solution $g_P(x)$ to Eq.~(\ref{eq:h-equation}) to arbitrary precision can be obtained by 
stabilising Eq.~(\ref{eq:gp-fixedpoint}) with a mixing strategy,
\begin{equation}
\begin{split}
g_P^{(0)}(x)=&\ h(x/P) \\
g_P^{(i+1)}(x)=&\ \alpha \left[h(x) -\sum_{k=1}^{P-1} \frac{g_P^{(i)}(x_k)}{x_k^2/x^2}\right] +(1-\alpha) g_P^{(i)}(x). 
\end{split}
\label{eq:gp-iterative}
\end{equation}
In particular, the mixing parameter $\alpha=1/P$ was found to give a sufficiently fast and 
convergent iteration for all of the numbers of beads we tried. 
The convergence of $g_P^{(i)}(x)$ to $g_P(x)$ is shown in Fig.~1 for $P=8$, along with examples 
of the resulting converged solutions for other values of $P$.  
Tabulated values of these solutions for $P$ ranging from 1 to 128 can be downloaded from
an on-line repository\cite{gle4md}. Once one has a converged $g_P(x)$, the frequency-dependent position 
fluctuation $c_{qq}(\omega)$ in Eq.~(\ref{eq:qp}) is given by
\begin{equation}
c_{qq}(\omega) = (Pk_BT/\omega^2)g_P(\hbar\omega/2k_BT), \label{eq:cqq-pigle}
\end{equation}
and the only remaining problem is to design a GLE that can be used to enforce this 
fluctuation on each ring polymer bead.

\subsection{Fitting and implementation}

The design of a GLE consistent with $c_{qq}(\omega)$ in Eq.~(\ref{eq:cqq-pigle}) consists 
of optimising the matrices in Eq.~(\ref{eq:mark-sde}) until the desired response of the 
thermostat is obtained. As has been found for a variety of other applications of the 
GLE,\cite{Ceriotti09a,Ceriotti09b,Ceriotti10a,Ceriotti10b} the efficacy of the resulting 
PI+GLE scheme will depend on the strategy by which the optimisation is performed, and on 
the additional criteria besides fitting $c_{qq}(\omega)$ that are used to define the merit 
function for the optimisation. For applications of the quantum thermostat 
to anharmonic problems the strength of the coupling between the thermostat and the Hamiltonian 
dynamics and the efficiency of the sampling can be just as important as the agreement 
between $c_{qq}(\omega)$ and the target function in Eq.~(18). 
The possibility of improving the accuracy systematically by increasing 
the number of beads makes zero-point energy leakage and other anharmonic effects a lesser concern 
in the present context. However, if systematic convergence is to be achieved, it is advisable that the sampling 
efficiency does not differ dramatically between the fits for different numbers of beads.

In the case of the original quantum thermostat described in Ref.~\cite{Ceriotti09b}, 
we chose to constrain $c_{pp}(\omega)=\omega^2c_{qq}(\omega)$ as in Eq.~(\ref{eq:cppcqq}). 
This had the advantage of providing ready access to the quantum mechanical momentum distribution, 
which is a quantity of direct relevance to recent deep inelastic neutron scattering experiments. 
However, by constraining $c_{pp}(\omega)$ in this way, we found that we had to enforce a strongly 
overdamped regime at low frequencies in order to avoid zero point energy leakage in applications 
to multi-dimensional systems.\cite{Ceriotti09b} In the present PI+GLE scheme, as in PIMD itself, 
the ring polymer momenta lose all physical significance as soon as there is more than one bead, 
and extracting the quantum mechanical momentum distribution requires a considerably more intricate 
calculation.\cite{Lin10} 
When it comes to fitting the GLE, it is therefore more natural to regard the momenta simply as 
a sampling device, and to focus exclusively on the fluctuations of configurations $c_{qq}(\omega)$, 
as we have already done in our description of PI+GLE in Sec.~III.C. 

As shown in Fig.~2, this leaves significantly more freedom in the fit, even in the case of just 
one bead. The extra freedom allows us to reproduce the desired $c_{qq}(\omega)$ with a maximum 
discrepancy smaller than $0.5$\% and to achieve high sampling efficiency over a broad range 
of frequencies, yielding a sampling performance comparable to that of an ``optimal sampling'' 
GLE.\cite{Ceriotti10a,Ceriotti10b} To give another example involving more beads, the self-diffusion 
coefficient for a model of liquid water (see Sec.~IV) - which in this context can only be regarded 
as a measure of the sampling efficiency for slow, collective motion - is reduced by less 
than 50\% in a well-converged (8 bead) PI+GLE simulation compared to NVE dynamics, whereas the original 
(one bead) quantum thermostat decreases the diffusion coefficient of the same model by a factor of 10. 

Having obtained a nearly-constant sampling efficiency is also beneficial to the transferability 
of the fitted parameters, as discussed in Refs.\cite{Ceriotti10a,Ceriotti10b}. 
As a matter of fact, the very same parameterization has been adopted for both of the examples given below, 
which are as different as a one-dimensional quartic double well and liquid water. 
With an appropriate scaling,\cite{Ceriotti09b,Ceriotti10a} 
these GLE parameters can be safely adopted in all circumstances where the maximum physical frequency 
present in the system is smaller than $35 k_B T/\hbar$. 
The GLE parameters we have used in the present study are available up to $P=16$ and may be 
downloaded from an online repository\cite{gle4md}.

The details of how we actually optimised the GLE matrices for the present PI+GLE application are 
rather technical, and less important than the criteria employed for the optimisation that we have 
just described. A comprehensive discussion of the optimisation of GLE matrices has recently been 
given elsewhere,\cite{Ceriotti10a} and we used exactly the same techniques in the present study. 
The implementation of a GLE thermostat into a PIMD code has also been discussed in detail in a 
recent paper,\cite{Ceriotti10b} where it was shown to be essentially no more complicated than 
adding a thermostat to a classical molecular dynamics simulation. However, in the case of PI+GLE 
there are a few additional points that we do need to make. 

First of all, the dynamics in PI+GLE {\em must} be performed using physical bead masses, as 
opposed to the alternative masses that are sometimes used in path integral schemes. 
Unless physical bead masses are used the frequencies of the necklace will be different from those 
in Eq.~(\ref{eq:omegak}), and Eq.~(\ref{eq:xk}) will have to be modified accordingly; 
this will change the functional equation for $g_P(x)$ in Eq.~(\ref{eq:h-equation}) and 
the new equation will have to be solved from scratch. The equations of motion can be 
integrated efficiently with physical bead masses by performing a normal mode 
transformation for the free ring polymer evolution\cite{Ceriotti10b} and/or employing a multiple time step scheme.\cite{Tuckerman92,Tuckerman93}

Another important observation is that, at variance with canonical sampling GLE schemes 
in which the free particle propagation of the GLE preserves the equilibrium distribution, 
the stationary probability distribution of the PI+GLE scheme requires an accurate integration 
of the stochastic differential equations of motion on the timescale of the fastest modes. 
For this reason, whenever a multiple time step integrator is used, the thermostat should 
be applied in the inner loop. As a consequence, the integration of the GLE introduces a 
sizeable computational overhead, which can be significant in cases when the calculation 
of physical forces is inexpensive. In the case of {\em ab initio} molecular dynamics, which 
is the primary target for the present method, the overhead will be completely negligible.

\section{The quartic double well\label{sec:dwell}}

As mentioned in the Introduction, one of the fundamental shortcomings of the original quantum 
thermostat\cite{Ceriotti09b} is its inability to provide a realistic description of tunnelling effects. 
The one-dimensional double well potential, 
\begin{equation}
V(q)=h \left[\left(\frac{2q}{d}\right)^2-1\right]^2,\label{eq:quartic}
\end{equation}
in which both zero-point energy and tunnelling are important, and can be tuned by adjusting the 
height $h$ of the barrier and the distance $d$ between the minima, therefore provides an ideal 
first example with which to benchmark the performance of our combined PI+GLE method.

In Fig.~3 we report the particle density $p(q)$ obtained in PIMD and PI+GLE simulations of this 
potential with $d=0.6$ \AA\ and $h=1000$ K at a temperature of 300 K, using different numbers of beads.
 By comparing the curves with the exact finite-temperature density 
\begin{equation}
\rho(q)=\sum_i e^{-\epsilon_i/k_BT} \left|\psi_i(q)\right|^2 /\sum_i e^{-\epsilon_i/k_BT},
\end{equation}
where $\epsilon_i$ and $\psi_i$ are the eigenvalues and eigenfunctions of the Schr\"odinger equation, 
one sees that the use of a GLE dramatically improves the results even when $P=1$. By increasing $P$ 
systematic convergence to exact result is achieved, and the convergence is greatly accelerated by the Langevin dynamics.

\begin{figure}[t]
\includegraphics{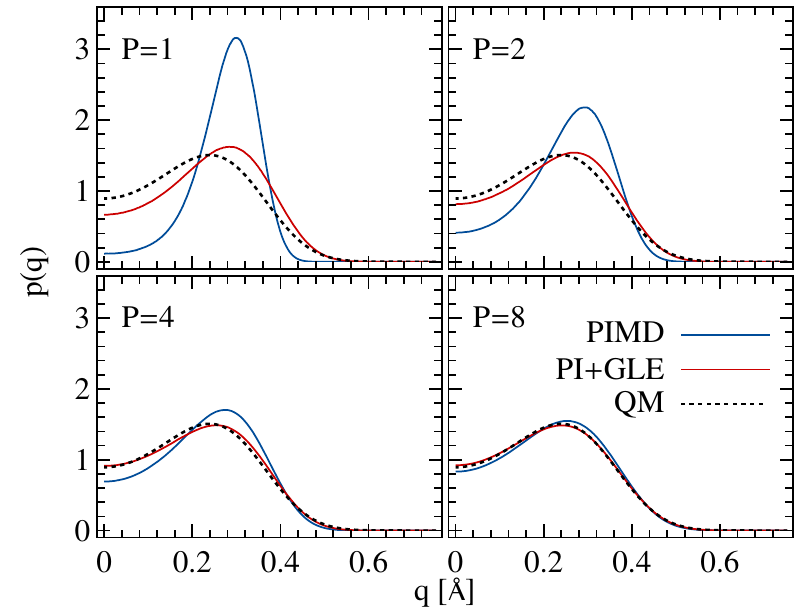}
\caption{Probability density for a hydrogen atom in a quartic double well potential with the minima separated by $0.6$~\AA\ and a barrier height of $1000$~K. All simulations were performed with a target temperature of $300$~K. The exact quantum mechanical result (dashed black line) was obtained by numerical solution of the Schr{\"o}dinger equation, the contributions of the various eigenstates being averaged with the appropriate Boltzmann weight. The four panels compare this exact result with the PIMD and PI+GLE results with increasing bead numbers.}
\end{figure}

In order to characterise quantitatively the convergence of the density, we computed the distance between 
$p(q)$ and $\rho(q)$, as a function of the parameters of the potential and the bead number. 
For the distance metric we employed the square root of the Jensen-Shannon divergence
\begin{equation}
d^2_{JS}(p,\rho)=\frac{1}{2}\int_{-\infty}^{\infty}
\left[p(q)\ln \frac{2 p(q)}{p(q)+\rho(q)}+\
\rho(q)\ln \frac{2\rho(q)}{p(q)+\rho(q)} \right]
dq, 
\end{equation}
which is a metric that is widely used to compare probability distributions.\cite{Endres03}

In Fig.~4 we present the convergence of $p(q)$ to $\rho(q)$ as the number of beads is increased. 
The simulations were performed using the mass of a hydrogen atom, different values of $d$ and $h$ as 
indicated in the figure, and a time step of $0.1$~fs. A very small time step was needed because 
we integrated the PI equations of motion directly using the velocity Verlet method, without 
exploiting the exact free ring polymer evolution that becomes possible in the normal mode representation.\cite{Ceriotti10b} 
For each set of parameters we ran PIMD and PI-GLE trajectories for $20$~ns. 

It is clear from Fig.~4 that the PI-GLE simulations yield a significantly better estimate of the finite-temperature 
density of the quartic double well than the PIMD simulations, even in the regime of small $d$ and 
large $h$ where tunnelling plays an important role. The addition of the GLE provides a given 
level of accuracy with an effort that is between two and eight times smaller than with a standard PI, 
depending on the parameters of the potential and on the accuracy required.

As the density approaches the exact $\rho(q)$, all physical observables that depend on the 
position of the particle converge to their quantum mechanical expectation values. 
In Fig.~5 we have plotted the average potential energy computed from the same trajectories 
that were used to compute the densities in Fig.~4. Again, the use of a GLE thermostat consistently improves the quality of results. 

\begin{figure}[t]
\includegraphics{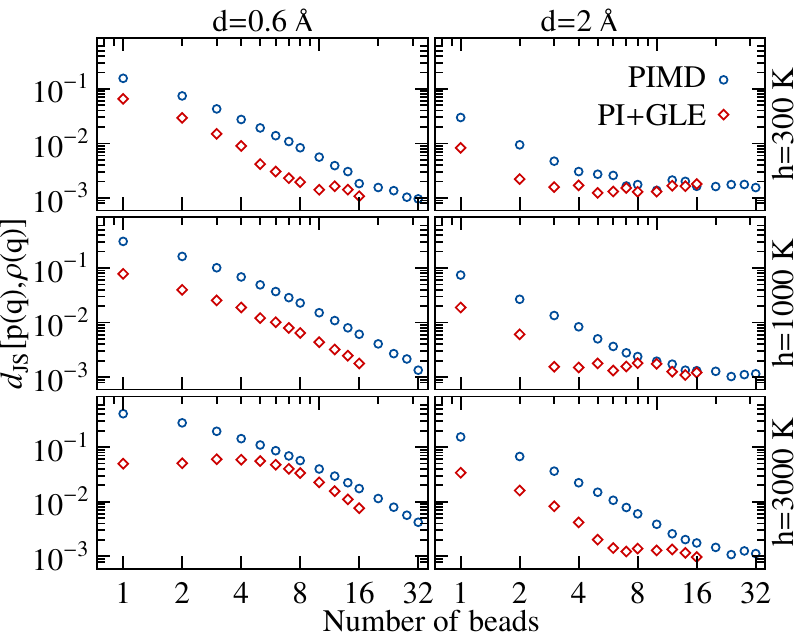}
\caption{Distance between the exact density $\rho(q)$ and the densities $p(q)$ obtained from PIMD and PI+GLE simulations with different bead numbers, for the quartic double well potential in Eq.~(\ref{eq:quartic}). Six different combinations of parameters of the double well potential have been tested. Note that in all cases the use of a GLE leads to a significant improvement in the estimate of the density. }
\end{figure} 

\begin{figure}[t]
\includegraphics{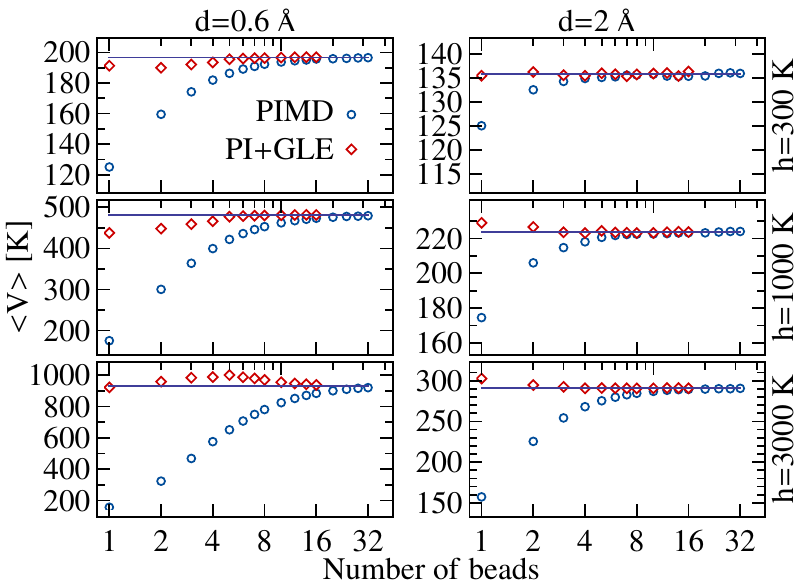}
\caption{Average potential energy for a hydrogen atom in a quartic double well potential computed by PIMD and PI+GLE, as a function of the number of beads and for different parameters of the potential. The exact, quantum mechanical expectation value is marked with a line. }
\end{figure}

\section{Liquid water}

In the previous section we have demonstrated, in a simple one-dimensional case, that an 
appropriately-constructed GLE can be used to accelerate the convergence of path integral 
molecular dynamics to the exact quantum mechanical probability distribution. 
However, as discussed in earlier publications,\cite{Ceriotti09b,Ceriotti10a} when applying 
the quantum thermostat to a multi-dimensional problem, one must be wary of zero point energy leakage. 
The GLE sets the various normal modes to different effective temperatures, and energy may flow from 
hot (high frequency) to cold (low frequency) modes because of anharmonic couplings. 

To assess the behavior of the combined PI+GLE strategy in this respect, and to evaluate the 
computational savings that can be expected from the GLE in a more typical application, we 
have performed some additional simulations for liquid water. For these simulations, we used 
a recently-developed flexible water model that was fit to reproduce a wide variety of properties 
of the liquid in path integral simulations.\cite{Habershon09a} 
This avoids the double counting of nuclear quantum effects that would result from the use 
of a force field fit to reproduce experimental data in classical dynamics. 
We performed simulations with both conventional PIMD and PI+GLE, using different numbers of beads. 
The refined ring polymer contraction scheme of Markland {\em et al.}\cite{Markland08c} was used 
to reduce the computational effort, with the long-range electrostatic interactions beyond 5~\AA\ 
contracted to the ring polymer centroid. The use of this scheme has been shown previously 
to provide significant computational savings in empirical force field simulations without 
affecting the accuracy of the results.\cite{Markland08c}

Simulations were performed with a target temperature $T=298$~K. We used a multiple-time step 
scheme with an outer time step of 0.75 fs and covalently bonded interactions computed every 
0.15 fs. We ran 3.15~ns trajectories for each set of parameters, with the first 150~ps used 
for equilibration. Ergodic constant-temperature sampling was achieved in the PIMD simulations 
by applying a targeted stochastic scheme to the internal modes of the necklace\cite{Ceriotti10b} 
and a global stochastic velocity rescaling to the centroid.\cite{Bussi07a}

In Fig.~6 we report the expectation values of the potential energy and the centroid virial 
kinetic energy, which were computed using standard path integral estimators.\cite{Ceperley95} 
In this case, the GLE is seen to be extremely useful, and leads to much faster convergence 
of averages compared to conventional PIMD. Interestingly, we found that to converge average 
energies to an error smaller than $1$ kJ/mol in the PIMD simulations, $P$ should be increased 
well beyond the $32$ beads that are generally adopted for room-temperature water. 
Conversely, PI+GLE yields results in perfect agreement with a $128$ bead PIMD simulation using 
fewer than $16$ beads.

\begin{figure}[t]
\includegraphics{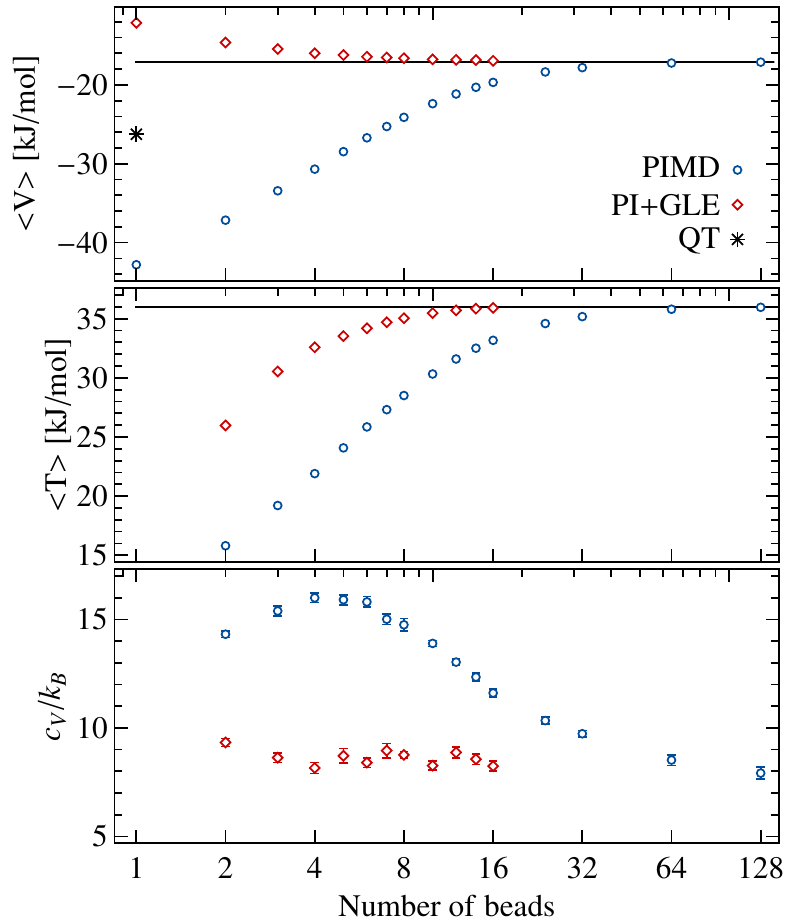}
\caption{The average value of the potential energy and the virial kinetic energy for a simulation of a flexible water model\cite{Habershon09a} at $T=298$~K, plotted as a function of the number of beads. The results obtained with conventional PIMD and PI+GLE are compared, and the value of $\left<V\right>$ obtained with the original quantum thermostat\cite{Ceriotti09b} (QT) is also reported. }
\end{figure}

\begin{figure*}[t]
\includegraphics{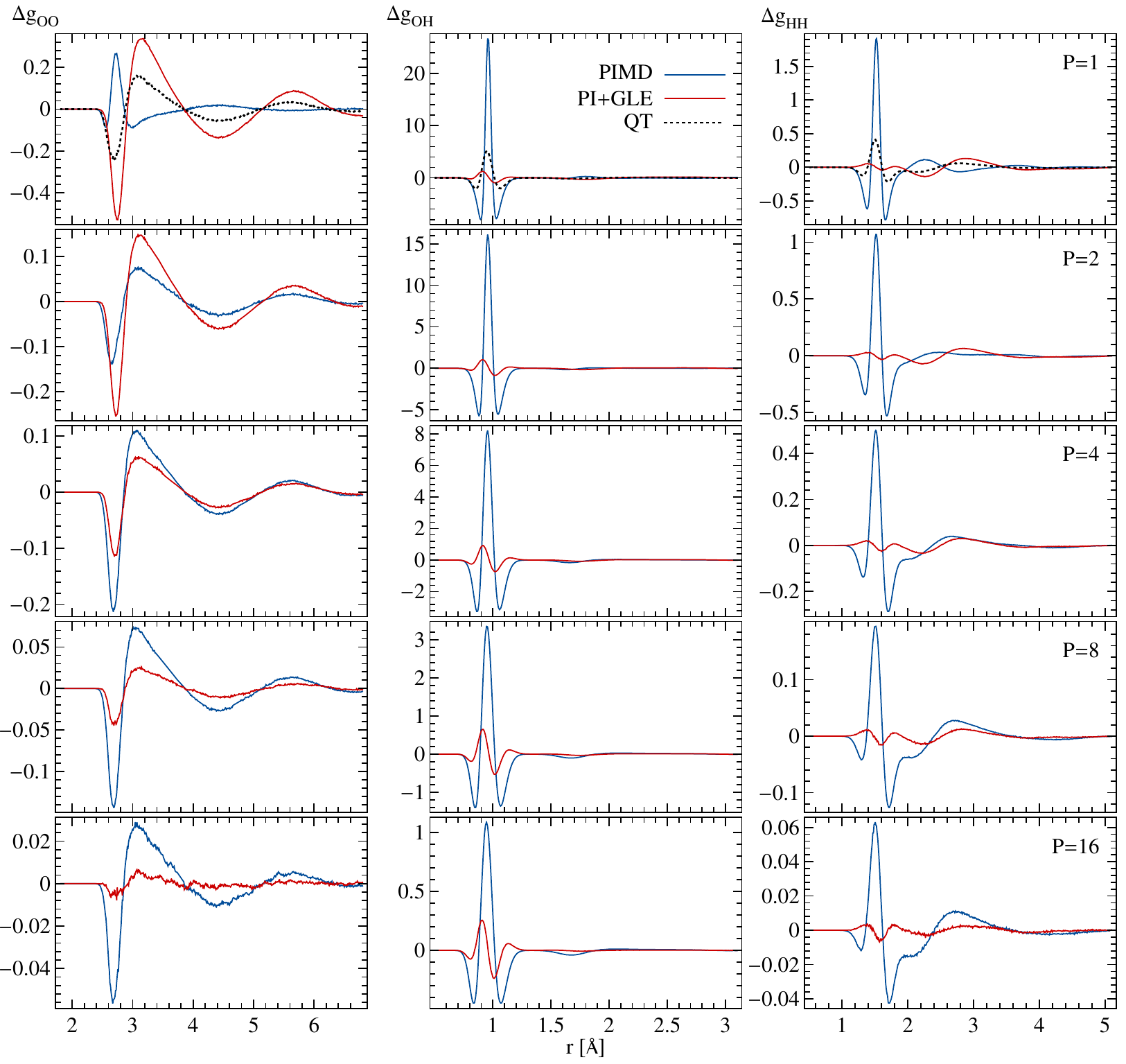}
\caption{Errors in the radial distribution functions of liquid water at $T=298$~K obtained in low-$P$ simulations using PIMD and PI+GLE, relative to a fully-converged PIMD reference simulation with 128 beads. Note that the scale of the $y$-axis is greatly magnified as $P$ is increased and the $\Delta g$'s become smaller. The results obtained using the original quantum thermostat\cite{Ceriotti09b} (QT) are also reported in the panels with $P=1$. The statistical error in the distribution functions is of the order of $10^{-3}$. }
\end{figure*} 

As a more sensitive benchmark of the convergence of the properties of quantum water we 
have also computed the constant-volume specific heat $c_V$,  which is known to require very 
large values of $P$ for an accurate determination.\cite{Shiga05} The value of and 
statistical uncertainty in $c_V$ were obtained from a quadratic fit 
to the values of $\left<V\right>$ and $\left<T\right>$ computed at $293$, $298$ and $303$~K. 
The PI+GLE method is again seen to perform exceedingly well in this test (see Fig.~6). 
The convergence is somewhat accelerated by a fortuitous cancellation between the errors in 
the PI+GLE estimates of the potential and kinetic energy contributions to~$c_V$, 
which are however both very well converged by the time $P\simeq 8$.

Comprehensive results for the convergence of structural properties of water are reported in 
Fig.~7, in which we systematically examine the radial distribution functions that are obtained 
with different bead numbers. Here again the GLE helps tremendously in accelerating the 
convergence of the PI calculation, in particular in the region of the intramolecular peaks,
 which are strongly affected by nuclear quantum effects. The only case in which the 
PI+GLE simulation is in worse agreement agreement with the exact result than the 
PIMD simulation is when the oxygen-oxygen $g(r)$ is computed with too few 
beads ($P=1$ and $P=2$). This is a consequence of the zero-point energy leakage in the 
PI+GLE simulation, which heats up low frequency degrees of freedom and washes out the 
long-range features from $g_{\rm OO}(r)$. By increasing the strength of the thermostat 
in the low-frequency region (as is done for instance in the case of the original 
quantum thermostat -- see Sec.~II.D) it is possible to mitigate this effect, at the 
expense however of a greater disturbance on diffusive modes, and hence on the 
efficiency of sampling. The PI+GLE strategy allows one to counter the effect of 
zero point energy leakage with a moderate increase in the number of beads, and indeed the 
PI+GLE radial distribution functions for liquid water are seen to be significantly more 
accurate than the PIMD radial distribution functions in Fig.~7 for all $P\ge 4$.

Combining the results in Figs.~6 and~7, one sees that PI+GLE provides the same accuracy 
in the thermodynamic and structural properties of liquid water with 8 beads as PIMD 
provides with 32 beads. This is the level of accuracy that is commonly accepted for 
room-temperature water simulations, and adding an appropriate GLE to the PI simulation 
allows it to be achieved with a four-fold reduction in the number of beads. 
The comparison becomes even more favourable if higher accuracy is required, with a 
16-bead PI+GLE simulation being as accurate as a 128-bead PIMD simulation for all 
of the observables we have considered.

\section{Conclusions}

In the present paper we have discussed and thoroughly demonstrated how a properly-designed 
generalized Langevin equation can be used to accelerate the convergence of path integral 
molecular dynamics to the exact quantum mechanical thermal expectation values. 
This leads to substantial savings in computational effort, the number of beads required to 
obtain a given level of accuracy being reduced by a factor of 4 or more. 
The original (one-bead) quantum thermostat\cite{Ceriotti09b} provides an even cheaper way 
to include nuclear quantum effects in mildly anharmonic systems, but it is 
an inherently approximate technique. The present PI+GLE combination captures tunnelling 
effects as well as zero point energy effects, and it can be systematically improved to 
give the exact quantum mechanical result simply by increasing the number of path integral beads. 
We expect that this combination will be particularly valuable when used in conjunction 
with {\em ab initio} molecular dynamics, in which the forces acting on the nuclei are so 
expensive to evaluate that nuclear quantum effects have only seldom been considered in the past. 

One final observation is that we have concentrated exclusively in this paper on the 
standard second order Trotter product path integral in Eqs.~(\ref{eq:pi-z}) and (\ref{eq:ham-p}). Another way to reduce 
the number of path integral beads is to use a higher-order imaginary time propagator, 
and some interesting progress has been made in this direction over the years.\cite{Takahashi84,Chin97,Jang01,Yamamoto05} 
This approach is fundamentally different from the approach we have investigated here, 
and potentially complementary to it. One could in principle develop a GLE scheme to 
accelerate the convergence of any imaginary time propagator, including the more promising 
of the higher-order propagators that have been suggested in the literature. 
In this way, one might hope to be able to reduce the number of beads even further, and 
make the inclusion of nuclear quantum effects in {\em ab initio} simulations almost routine. 
In any event, the results we have presented in this paper clearly demonstrate the potential 
of the generalized Langevin equation as a computational tool for accelerating the convergence of PIMD.

\section{Acknowledgments}
We would like to thank Oliver Riordan for a helpful discussion about the functional 
equation for $g_P(x)$ in Eq.~(\ref{eq:h-equation}) and the non-uniqueness of its solutions, 
and Giovanni Bussi and Thomas Markland for stimulating and insightful conversations.

\end{document}